\documentclass[epj,nopacs]{svjour}
\usepackage{graphics}
\usepackage{epsfig}
\def\Journal#1#2#3#4{{#1}{\bf #2}, (#4) #3 }
\def\IJMPA{{Int. J. Mod. Phys. A}}
\def\JPG{{J. Phys.}~{\bf G}}

\def\NPB{{Nucl. Phys.}~{\bf B}}
\def\PLB{{Phys. Lett. B}}
\def\PLC{Phys. Repts.\ }
\def\PRL{Phys. Rev. Lett.\ }
\def\PRD{{Phys. Rev. D}}
\def\PRC{{Phys. Rev.  C}}

\def\ARNS{{Ann. Rev. Nucl. Part. Sci.\ }} 
\def\RMP{Rev. Mod. Phys.\ }

\begin{document}
\title{Hard scattering and jets---from p-p collisions in the 1970's to Au+Au collisions at RHIC.}
\author{M.~J.~Tannenbaum 
\thanks{Research supported by U.S. Department of Energy, DE-AC02-98CH10886.}
}                     
%
%
\institute{Brookhaven National Laboratory\\Upton, NY 11973-5000 USA}
\date{Received: date / Revised version: date}
%
\abstract{Hard scattering in p-p collisions, discovered at the CERN 
ISR in 1972 by the method of leading particles, proved that the partons of Deeply Inelastic Scattering strongly interacted with each other. Further ISR measurements utilizing inclusive 
single or pairs of hadrons established that high 
$p_T$ particles are produced from states with two roughly back-to-back jets which are 
the result of scattering of constituents of 
the nucleons as described by Quantum Chromodynamics (QCD), which was developed during the course of these measurements.  
These techniques, which are the only practical method to study hard-scattering and jet 
phenomena in Au+Au central collisions, are reviewed, with application to measurements at RHIC.}
%
%
\authorrunning{M.~J.~Tannenbaum}
\titlerunning{Hard scattering---from p-p collisions in the 1970's to Au+Au collisions at RHIC}
\maketitle
\section{Introduction}
\label{intro}
  In 1998, at the QCD workshop in Paris, Rolf Baier asked me whether jets could be measured in Au+Au collisions because he had a prediction of a QCD medium-effect on color-charged partons traversing a hot-dense-medium composed of screened color-charges~\cite{BaierQCD98}. I told him~\cite{MJTQCD98} that there was a general consensus~\cite{Strasbourg} that for Au+Au central collisions at $\sqrt{s_{NN}}=200$ GeV, leading particles are the only way to find jets, because in one unit of the nominal jet-finding cone,  $\Delta r=\sqrt{(\Delta\eta)^2 + (\Delta\phi)^2}$, there is an estimated $\pi\times{1\over {2\pi}} {dE_T\over{d\eta}}\sim 375$ GeV of energy !(!)
  
	The good news was that hard-scattering in p-p collisions had been discovered at the CERN 1SR~\cite{CCR,SS,BS} by the method of leading particles, before the advent of QCD, and it was proved by single inclusive and two-particle correlation measurements in the period 1972-1982 that high $p_T$ particles are produced from states with two roughly back-to-back jets which are the result of scattering of constituents of the nucleons as described by QCD, which was developed during this period. The other good news was that the PHENIX detector had been designed to make such measurements and could identify and separate direct single $\gamma$ and $\pi^0$ out to $p_T\geq 30$ GeV/c.

\section{Systematics of single particle inclusive production in p-p collisions.}
\label{sec:single}
\begin{figure}[!h]
\vspace*{-0.25in}
\includegraphics[width=0.45\textwidth]{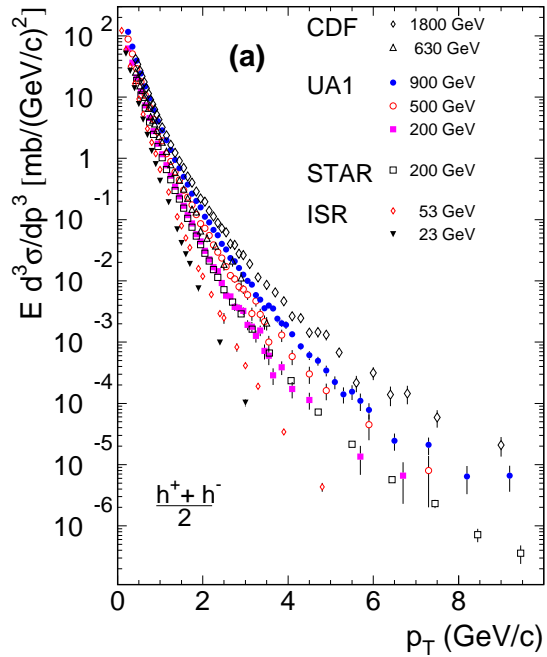} 
\vspace*{-0.35in}
\caption[]{$E {d^3\sigma}/{d^3p}$ vs. $p_T$ at mid-rapidity as a function of
$\sqrt{s}$ in $p-p$ collisions.}
\label{fig:hpT}
\end{figure}

In p-p collisions, the invariant cross section for non identified 
charge-averaged hadron production at 90$^\circ$ in the c.m. system as a 
function of the transverse 
momentum $p_T$ and c.m. energy $\sqrt{s}$ has a characteristic 
shape (Fig.~\ref{fig:hpT}). There is an 
exponential tail ($e^{-6p_T}$) at low $p_T$, which depends very little on 
$\sqrt{s}$. This is the soft physics region, 
where the hadrons are fragments of 
`beam jets'. At higher $p_T$, there is a power-law tail which depends very 
strongly on $\sqrt{s}$. 
This is the hard-scattering region, where the hadrons 
are fragments of the high $p_T$ QCD jets from constituent-scattering. 

	The hard scattering behavior for the reaction $p+p\rightarrow C+X$ is easy to understand from general principles proposed by Bjorken and collaborators~\cite{Bj,BBK} and subsequent authors~\cite{CIM,CGKS}.  Using the principle of factorization of the reaction into parton distribution functions for the protons, fragmentation functions to particle $C$ for the scattered partons and a short-distance parton-parton hard scattering cross section, the invariant cross section for the inclusive reaction, where particle $C$ has transverse momentum $p_T$ near mid-rapidity, was given by the general `$x_T$-scaling' form~\cite{CIM}, where $x_T=2p_T/\sqrt{s}$: 
\begin{equation}
E \frac{d^3\sigma}{dp^3}=\frac{1}{p_T^{n}} F({2 p_T \over \sqrt{s}})
= \frac{1}{\sqrt{s}^{\,n}} G({x_T}) .
\label{eq:bbg}
\end{equation}
The cross section has 2 factors, a function $F$ $(G)$ which `scales', i.e. depends only on the ratio of momenta; and a dimensioned factor, ${p_T^{-n}}$   $(\sqrt{s}^{\,-n})$,   
where $n$ gives the form of the force-law 
between constituents. For QED or Vector Gluon exchange~\cite{BBK}, $n=4$, and for the case of quark-meson scattering by the exchange of a quark~\cite{CIM}, $n$=8. When QCD is added to the mix~\cite{CGKS}, pure scaling breaks down and $n$ varies according to the $x_T$ and $\sqrt{s}$ regions used in the comparison, $n\rightarrow n(x_T, \sqrt{s})$. 
\begin{figure}[!hbt]
\vspace*{-0.25in}
\includegraphics[width=0.45\textwidth]{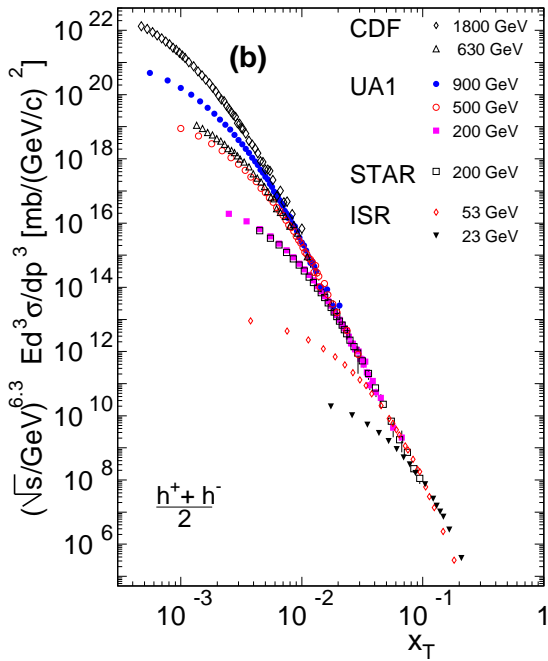}
\vspace*{-0.25in}
\caption[]{$\sqrt{s}({\rm GeV})^{6.3}\times Ed^3\sigma/d^3p$ vs $x_T=2{p_T}/\sqrt{s}$. }
\label{fig:hxT}
\end{figure}

	We now know that the characteristic $\sqrt{s}$ dependence of the high $p_T$ tail is simply explained by the $x_T$ scaling of the spectrum (with $n=6.3$, valid in the range $0.01\leq x_T\leq 0.1$ relevant to the early RHIC measurements (see Fig.~\ref{fig:hxT}))~\cite{ppg023}. However, it is worthwhile to note that it took quite some time for $x_T$ scaling with the value of $n=5.1\pm 0.4$, consistent with QCD, to be observed at the CERN-ISR~\cite{CCOR}. This was due to the so-called `intrinsic' transverse momentum of partons, the ``$k_T$ effect'', which causes a transverse momentum imbalance of the outgoing parton-pairs from hard-scattering, making the jets not exactly back-to-back in azimuth. This was discovered by experimenters~\cite{CCHK} and clarified by Feynman and collaborators~\cite{FFF}. The ``$k_T$-effect'' acts to broaden the $p_T$ spectrum, thus spoiling the $x_T$-scaling at values of $p_T\leq 7.5$ GeV/c, at the ISR, and totally confusing the issue at fixed target incident energies of 200--400 GeV~\cite{Cronin,MJT79} due to the the relatively steep $p_T$ spectrum (see Fig.~\ref{fig:hpT}), which results in a relatively strong broadening effect. It is also evident from Fig.~\ref{fig:hpT} that hard-scattering, which is a relatively small component of the $p_T$ spectrum at $\sqrt{s}\sim 20$ GeV, dominates for $p_T\geq 2$ GeV/c by nearly 2 orders of magnitude at RHIC c.m. energies compared to the soft physics $e^{-6p_T}$  extrapolation~\cite{ppg031}.     

    The status of theory and experiment, circa 1980, is summarized by the first modern QCD calculation and prediction for high $p_T$ single particle production in hadron-hadron collisions,  in agreement with the data. The calculation by Jeff Owens and collaborators~\cite{Owens78} included  non-scaling and initial state radiation  under the assumption that high $p_T$ particles  are produced from states with two roughly back-to-back jets which are the result of scattering of constituents of the nucleons (partons). 
   The overall $p+p$ hard-scattering cross section in ``leading logarithm" pQCD   
is the sum over parton reactions $a+b\rightarrow c +d$ 
(e.g. $g+q\rightarrow g+q$) at parton-parton center-of-mass (c.m.) energy $\sqrt{\hat{s}}=\sqrt{x_1 x_2 s}$.  
\begin{equation}
\frac{d^3\sigma}{dx_1 dx_2 d\cos\theta^*}=
\frac{1}{s}\sum_{ab} f_a(x_1) f_b(x_2) 
\frac{\pi\alpha_s^2(Q^2)}{2x_1 x_2} \Sigma^{ab}(\cos\theta^*)
\label{eq:QCDabscat}
\end{equation} 
where $f_a(x_1)$, $f_b(x_2)$, are parton distribution functions, 
the differential probabilities for partons
$a$ and $b$ to carry momentum fractions $x_1$ and $x_2$ of their respective 
protons (e.g. $u(x_2)$), and where $\theta^*$ is the scattering angle in the parton-parton c.m. system. The characteristic subprocess angular distributions,
{\bf $\Sigma^{ab}(\cos\theta^*)$},
and the coupling constant,
$\alpha_s(Q^2)=\frac{12\pi}{25} \ln(Q^2/\Lambda^2)$,
are fundamental predictions of QCD~\cite{CutlerSivers,Combridge:1977dm}.

\begin{figure*}[ht]
\begin{center}
\begin{tabular}{cc}
\hspace*{-0.1in}\includegraphics[width=0.75\linewidth]{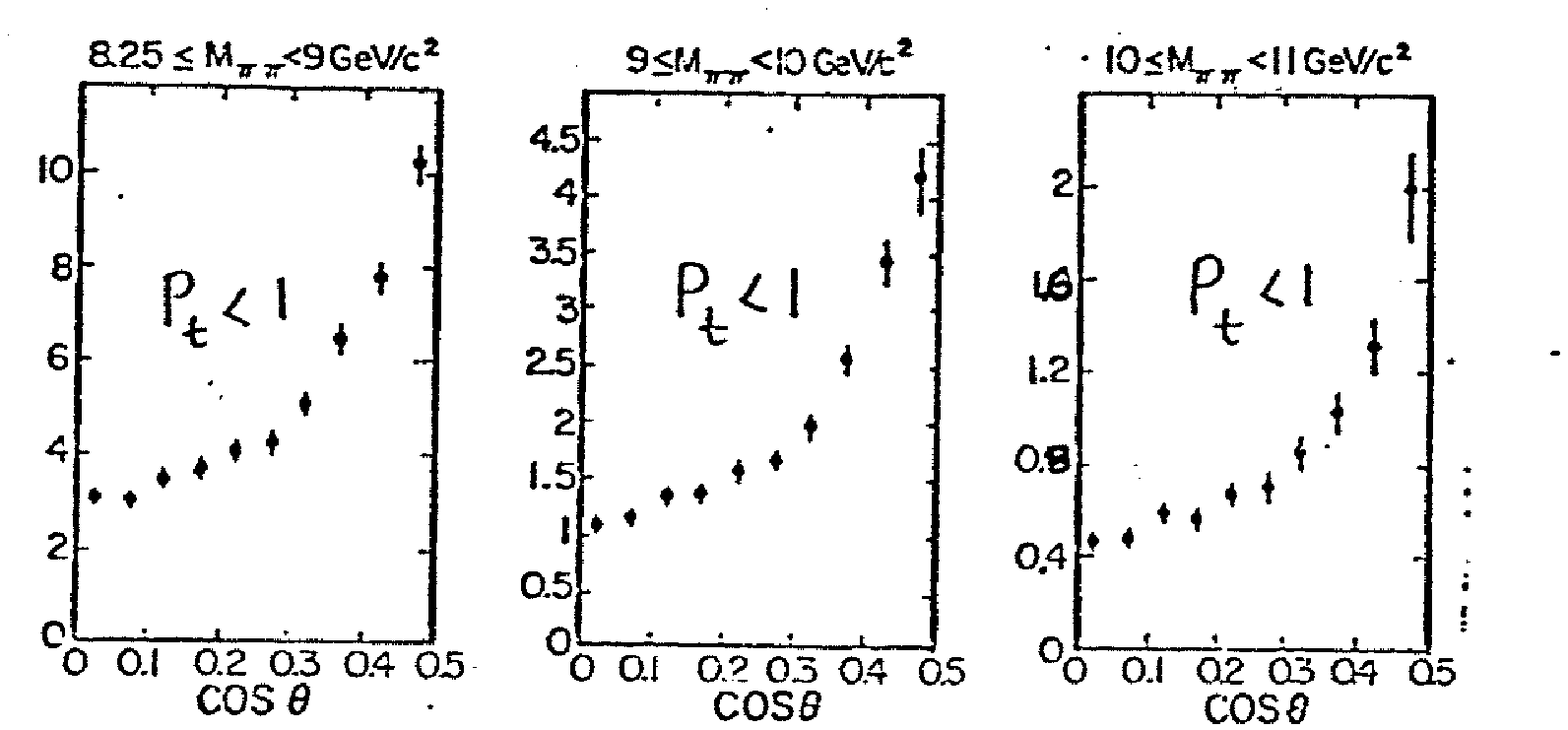} &
\hspace*{-0.39in}\includegraphics[width=0.288\linewidth]{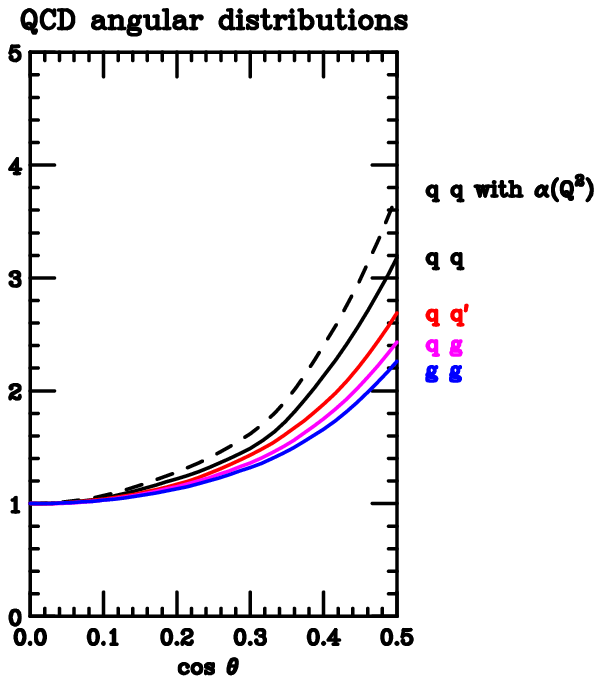}
\end{tabular}
\end{center}
\caption[]
{a) (left 3 panels) CCOR measurement~\cite{Paris82,CCOR82NPB} of polar angular distributions of $\pi^0$ pairs with net $p_T < 1$ GeV/c at mid-rapidity in p-p collisions with $\sqrt{s}=62.4$ GeV for 3 different values of $\pi\pi$ invariant mass $M_{\pi \pi}$. b) (rightmost panel) QCD predictions for $\Sigma^{ab}(\cos\theta^*)$ for the elastic scattering of $gg$, $qg$, $qq'$, $qq$, and $qq$ with $\alpha_s(Q^2)$ evolution.    
\label{fig:ccorqq} }
\end{figure*}

	The difficulty in finding jets in $4\pi$ calorimeters at ISR energies or lower gave rise to many false claims, creating skepticism during the period 1977-82~\cite{MJTIJMPA}, although jet effects are simply and directly visible using 2-particle correlations of high $p_T$ particles. 
A `phase change' in belief-in-jets was produced by one UA2 event 
at the 1982 ICHEP in Paris~\cite{Paris82}, which, together with the first direct measurement of the QCD constituent-scattering angular distribution, $\Sigma^{ab}(\cos\theta^*)$ (Eq.~\ref{eq:QCDabscat}), using two-particle correlations, presented at the same meeting (Fig.~\ref{fig:ccorqq}), gave universal credibility to the pQCD description of high $p_T$ hadron physics~\cite{Owens,Darriulat,DiLella}. The measurement of jets and jet properties via 2-particle correlations was a key element in understanding the details of high $p_T$ production. 

\section{Almost everything you want to know about jets can be found with 2-particle correlations.} 
\label{sec:pair}
   Many ISR experiments provided excellent 2-particle correlation measurements~\cite{Moriond79}. The CCOR experiment~\cite{Angelis79} was the first
 \begin{figure}[ht]
\vspace*{-0.25in}
\begin{center}
\includegraphics[width=0.50\textwidth]{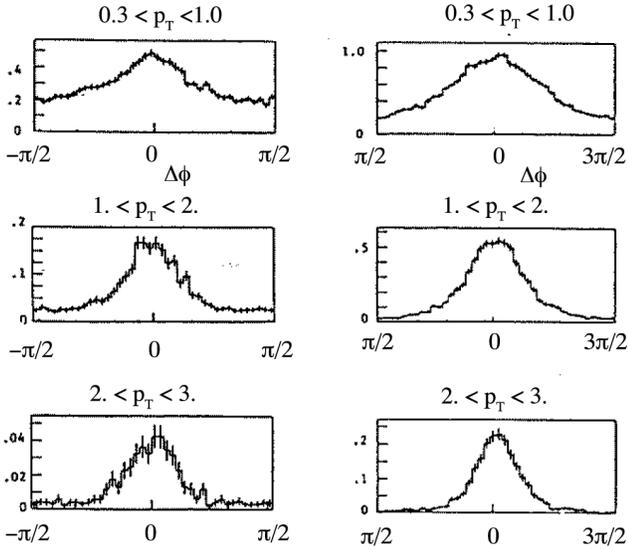} 
\end{center}
\caption[]
{a,b) Azimuthal distributions of charged particles of transverse momentum $p_T$, with respect to a trigger $\pi^0$ with $p_{Tt}\geq 7$ GeV/c, for 3 intervals of $p_T$: a) for $\Delta\phi=\pm \pi/2$ rad about the trigger particle, and b) for $\Delta\phi=\pm \pi/2$ about $\pi$ radians (i.e. directly opposite in azimuth) to the trigger. The trigger particle is restricted to $|\eta|<0.4$, while the associated charged particles are in the range $|\eta|\leq 0.7$.   
\label{fig:ccorazi} }
\end{figure}
to provide charged particle measurement with full and uniform acceptance over the entire azimuth, with pseudorapidity coverage $-0.7\leq \eta\leq 0.7$, so that the jet structure of high $p_T$ scattering could be easily seen and measured. In  Fig.~\ref{fig:ccorazi}a,b, the azimuthal distributions of associated charged particles 
relative to a $\pi^0$ trigger with transverse momentum $p_{Tt} > 7$ GeV/c are shown for three intervals of associated particle transverse momentum $p_T$. In all cases, strong correlation peaks on flat backgrounds are clearly visible, indicating the di-jet structure which is contained in an interval $\Delta\phi=\pm 60^\circ$ about a direction towards and opposite the to trigger for all values of associated $p_T\, (>0.3$ GeV/c) shown. The width of the peaks about the trigger direction (Fig.~\ref{fig:ccorazi}a), or opposite to the trigger (Fig.~\ref{fig:ccorazi}b) indicates out-of-plane activity from the individual fragments of jets. The fact that the width ($\Delta\phi$) of the away peak (Fig.~\ref{fig:ccorazi}b) does not decrease in proportion to $\sim \langle j_T\rangle /p_T$, where $\langle j_T\rangle$ is the mean transverse momentum of jet fragmentation, is indicative of the fact that the angular width of the away peak is dominated by the jet acoplanarity due to $k_T$, and not by the transverse momentum of fragmentation, $j_T$. 

\begin{figure}[ht]
\includegraphics[width=0.450\textwidth,angle=-1]{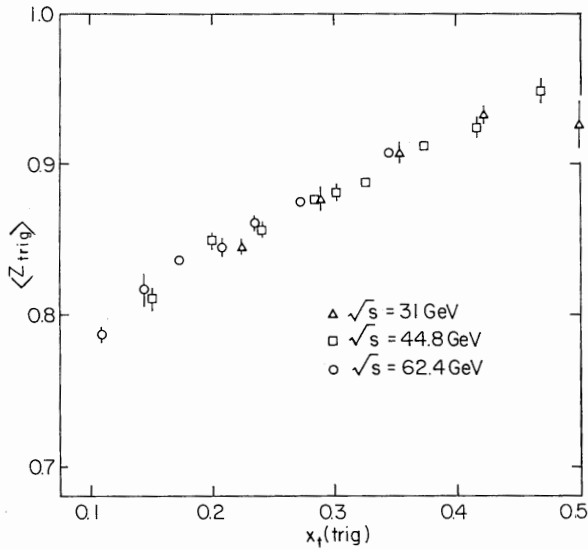}
\caption[]
{CCOR~\cite{CCOR82NPB} measurement of $\langle z_{\rm trig}\rangle$ as a function of $x_{Tt}=2p_{Tt}/\sqrt{s}$.  
\label{fig:ccormeanz} }
\end{figure}

	The same side peak shows the important property of ``trigger bias''~\cite{JacobLandshoff}  on which the method of leading particles is based: due to the steeply falling power-law transverse momentum spectrum of the scattered partons, the inclusive single particle (e.g. $\pi$) spectrum from jet fragmentation is dominated by fragments with large $z$, where $z=p_{T\pi}/p_{T_q}$ is the fragmentation variable. The trigger bias was directly measured from these data by reconstructing the trigger jet from the associated charged particles with $p_T\geq 0.3$ Gev/c, within $\Delta\phi=\pm 60^\circ$ from the trigger particle, using the algorithm $p_{T{\rm jet}}=p_{Tt}+1.5\sum p_T\cos(\Delta\phi)$, where the factor 1.5 corrects the measured charged particles for missing neutrals. The measured $z_{\rm trig}=p_{Tt}/p_{T{\rm jet}}$ distributions for 3 values of $\sqrt{s}$ (Fig.~\ref{fig:ccormeanz}) show the ``unexpected''~\cite{JacobEPS79} property of $x_T$ scaling. The jet properties,  $j_T$ and $k_T$ were also measured from these data~\cite{CCOR80}, with the result that $\langle j_T\rangle$ is  a constant, independent of  $p_{T_t}$ and $\sqrt{s}$, as expected for fragmentation, but $k_T$ increases with both $p_{T_t}$ and $\sqrt{s}$, suggestive of a radiative origin, rather than an `intrinsic' origin due to confinement.  

\section{Application to RHIC}
In 1998~\cite{MJTQCD98}, inspired by Rolf and collaborators, and before them by the work of Gyulassy~\cite{MGyulassy} and Wang~\cite{XNWang}, I indicated that my best bet on discovering the QGP was to 
utilize semi-inclusive $\pi^0$ or $\pi^{\pm}$ production. I expressed my hope that the QGP would cause the hard-scattered, high $p_T$ partons to lose all their energy and stop, so that the high $p_T$ tail would `vanish' for central Au+Au collisions. If the power-law tail would 
return when peripheral Au+Au collisions are selected, then this would be 
proof of a hot/dense/colorful medium (QGP??) in central Au+Au collisions. 
This is apparently what we see at RHIC~\cite{PXsuppression}.  

	The results of sections~\ref{intro}--\ref{sec:pair} enabled us to understand that $\pi^0$ with $p_T\geq 2$ GeV/c at mid-rapidity are produced, at RHIC, by hard-scattering in the region of $x$ where QCD with $T_{AB}$-scaled structure functions is valid, so that the huge suppression of $\pi^0$ observed in central Au+Au collisions was indisputably new physics. These same simple arguments revealed that the behavior of the $p$ and $\overline{p}$ in the range $2\leq p_T\leq 4.5$ GeV/c in Au+Au collisions was anomalous, another important discovery~\cite{PXpbar} which was totally unanticipated and is as yet unexplained. 
	
	It is rewarding to see that the methods and concepts discussed here, such as $j_T$, $k_T$~\cite{JRakQM}, $x_T$ scaling~\cite{ppg023} and 2-particle correlations~\cite{STAR},  are now in common use at RHIC as tools for gaining an understanding of the basic physics of jet suppression and its use  as a probe of the medium produced.

%

\begin{thebibliography}{}
%
%
\bibitem{BaierQCD98} R.~Baier, {\em QCD, Proc. IV Workshop}, 
eds. H.~M.~Fried and B.~M\"uller (World Scientific, Singapore, 1999), 
pp 272--279.   
\bibitem{MJTQCD98}M.~J.~Tannenbaum, {\it ibid.}, 
pp 280--285, pp 312--319.
\bibitem{Strasbourg} e.g. see {\em Proc. Int'l Wks. Quark Gluon Plsama 
Signatures}, eds. V.~Bernard, {et al.}, (Editions Frontieres, Gif-sur-Yvette, 
France, 1991).   
\bibitem{CCR} F.~W.~B\"usser, {\it et al.}, \Journal{\PLB}{46}{471}{1973};  
see also {Proc. 16th Int. Conf. HEP}, eds. J.~D.~Jackson and A.~Roberts, 
(NAL, Batavia, IL, 1972) Vol.~3, p.~317. 
\bibitem{SS} M.~Banner, {\it et al.}, \Journal{\PLB}{44}{537}{1973}. 
\bibitem{BS} B.~Alper, {\it et al.}, \Journal{\PLB}{44}{521}{1973}. 
\bibitem{Bj} J.~D.~Bjorken, \Journal{\PRD}{179}{1547}{1969}.  
\bibitem{BBK} S.~M.~Berman, J.~D.~Bjorken and J.~B.~Kogut, 
\Journal{\PRD}{4}{3388}{1971}.
\bibitem{CIM} R.~Blankenbecler, S.~J.~Brodsky, J.~F.~Gunion, 
\Journal{\PLB}{42}{461}{1972}. 
\bibitem{CGKS} R.~F.~Cahalan, K.~A.~Geer, J.~Kogut and Leonard Susskind, 
\Journal{\PRD}{11}{1199}{1975}.
\bibitem{ppg023} S.~S.~Adler, {\it et al.}, \Journal{\PRC}{69}{034910}{2004}.
\bibitem{CCOR} A.~L.~S.~Angelis, {\it et al.}, 
\Journal{\PLB}{79}{505}{1978}. 
See also, A.~G.~Clark, {\it et al.}, 
\Journal{\PLB}{74}{267}{1978}.  
\bibitem{CCHK} M.~Della Negra, {\it et al.}, 
\Journal{\NPB}{127}{1}{1977}. 
\bibitem{FFF} R.~P.~Feynman, R.~D.~Field and G.~C.~.Fox, \Journal{\NPB}{128}{1}{1977}.
\bibitem{Cronin} D.~Antreasyan, J.~W.~Cronin, {\it et al.}, 
\Journal{\PRL}{38}{112}{1977}. 
\bibitem{MJT79} For a contemporary view of the excitement of this period, and some more details, see M.~J.~Tannenbaum, Particles and Fields-1979, AIP Conference Proceedings Number 59, eds. B.~Margolis, D.~G.~Stairs, (American Institute of Physics, New York, 1980) pp. 263-309.
\bibitem{ppg031} S.~S.~Adler, {\it et al.}, \Journal{\PRL}{93}{202002}{2004}.
\bibitem{Owens78} J.~F.~Owens, E.~Reya, M.~Gl\"uck, 
\Journal{\PRD}{18}{1501}{1978}; J.~F.~Owens and J.~D.~Kimel, 
\Journal{\PRD}{18}{3313}{1978}.  
\bibitem{CutlerSivers} R.~Cutler and D.~Sivers, \Journal{\PRD}{17}{196}{1978}; \Journal{\PRD}{16}{679}{1977}.
\bibitem{Combridge:1977dm} B.~L.~Combridge, J.~Kripfganz and J.~Ranft, \Journal{\PLB}{70}{234}{1077}.
\bibitem{MJTIJMPA} e.g. for a review, see M.~J.~Tannenbaum, 
\Journal{\IJMPA}{4}{3377}{1989}. 
\bibitem{Paris82} { Proc. 21st Int'l Conf. HEP}, Paris, 1982, eds 
P.~Petiau, M.~Porneuf, J. Phys. C{\bf 3}\ (1982): see J.~P.~Repellin, p.  
C3-571; also see M.~J.~Tannenbaum, p. C3-134, G.~Wolf, p. C3-525.  
\bibitem{Owens} J.~F.~Owens, \Journal{\RMP}{59}{465}{1987}.
\bibitem{Darriulat} P.~Darriulat,\Journal{\ARNS}{30}{159}{1980}.
\bibitem{DiLella} L.~DiLella, \Journal{\ARNS}{35}{107}{1985}.
\bibitem{CCOR82NPB} A.~L.~S.~Angelis, {\it et al.}, 
\Journal{\NPB}{209}{284}{1982}.
\bibitem{Moriond79} e.g. see Proc. XIV Rencontre de Moriond, March 11-23, 1979, Les Arcs, France, ``Quarks, Gluons and Jets", ed. J. Tran Thanh Van (Editions Fronti\`eres, Dreux, France, 1979), H.~Boggild, p. 321, M.~J.~Tannenbaum, p. 351, and references therein.  

\bibitem{Angelis79} A.~L.~S.~Angelis, {\it et al.}, Physica Scripta 19\ (1979) 116.
\bibitem{JacobLandshoff} M.~Jacob and P.~Landshoff, \Journal{\PLC}{48}{286}{1978}. 
\bibitem{JacobEPS79} M.~Jacob, Proc. EPS International Conference on High-Energy Physics, Geneva, 27\ June-4\ July 1979 (CERN, Geneva, 1979) Volume 2, pp. 473-522. 
\bibitem{CCOR80} A.~L.~S.~Angelis, {\it et al.}, \Journal{\PLB}{97}{163}{1980}. 
\bibitem{MGyulassy} Miklos Gyulassy and Michael Pl\"umer, \Journal{\PLB}{243}{432}{1990}. 
\bibitem{XNWang} Xin-Nian Wang and Miklos Gyulassy, \Journal{\PRL}{68}{1480}{1992}. 
\bibitem{PXsuppression} K.~Adcox, {\it et al.}, \Journal{\PRL}{88}{022301}{2002}; S.~S.~Adler,    
{\it et al.}, \Journal{\PRL}{91}{072301}{2003}.
\bibitem{PXpbar} K.~Adcox, {\it et al.}, \Journal{\PRL}{88}{242301}{2002}.
\bibitem{JRakQM} J.~Rak, {\it et al.}, \Journal{\JPG}{30}{S1309}{2004}.
\bibitem{STAR} C.~Adler, {\it et al.}, \Journal{\PRL}{90}{082302}{2003}.  
\end{thebibliography}
%

\end{document}